# Isambard-AI: a leadership class supercomputer optimised specifically for Artificial Intelligence

Simon McIntosh-Smith
University of Bristol
Bristol, United Kingdom
S.McIntosh-Smith@bristol.ac.uk

Sadaf R Alam
University of Bristol
Bristol, United Kingdom
Sadaf.Alam@bristol.ac.uk

Christopher Woods
University of Bristol
Bristol, United Kingdom
Christopher.Woods@bristol.ac.uk

*Abstract* — **Isambard-AI is a new, leadership-class supercomputer, designed to support AI-related research. Based on the HPE Cray EX4000 system, and housed in a new, energy efficient Modular Data Centre in Bristol, UK, Isambard-AI employs 5,448 NVIDIA Grace-Hopper GPUs to deliver over 21 ExaFLOP/s of 8-bit floating point performance for LLM training, and over 250 PetaFLOP/s of 64-bit performance, for under 5MW. Isambard-AI integrates two, all-flash storage systems: a 20 PiByte Cray ClusterStor and a 3.5 PiByte VAST solution. Combined these give Isambard-AI flexibility for training, inference and secure data accesses and sharing. But it is the software stack where Isambard-AI will be most different from traditional HPC systems. Isambard-AI is designed to support users who may have been using GPUs in the cloud, and so access will more typically be via Jupyter notebooks, MLOps, or other web-based, interactive interfaces, rather than the approach used on traditional supercomputers of ssh'ing into a system before submitting jobs to a batch scheduler. Its stack is designed to be quickly and regularly upgraded to keep pace with the rapid evolution of AI software, with full support for containers. Phase 1 of Isambard-AI is due online in May/June 2024, with the full system expected in production by the end of the year.**

*Keywords—Artificial Intelligence (AI), Large Language Models (LLMs), Supercomputing, Modular Data Centre, GPU accelerated computing, AI safety and trustworthiness.*

I. INTRODUCTION

Isambard-AI is a new AI Research Resource, or AIRR, for the UK to quickly address the lack of national AI-capable supercomputing facilities for open research, and the impact this was having on AI research, as well as AI-enabled scientific discovery. The system was announced prior to the international AI Safety summit on November 1-2 2023, that was led by the UK's prime minister [1]. Hence, AI safety and trustworthiness research is a priority among the mission areas for AI RR [2]. The main challenge for the Isambard team was to build this system from zero (no data centre) to an operational facility in less than a year. This manuscript outlines that architecture and our design choices for developing a sustainable and performant solution in record time. Isambard-AI has achieved a number of significant firsts, including: the first time a high-density, HPE EX Direct Liquid Cooled solution has been integrated into a Modular Data Centre (MDC); first hybrid MDC solution including both air cooled (Grace-Grace) and direct liquid cooled (Grace-Hopper) systems in the same MDC for AI and HPC workloads; fastest installation for an end-end data centre and supercomputing solution to be deployed, from contract signature, design, build, order, integration, delivery to operational lights on inside 4 months. The early results from the Isambard-AI phase 1 system validate our design choices, from the data centre to direct liquid cooled cabinets to the AI optimised compute ecosystem.

At the time of writing this manuscript in the first week of April'24, we have demonstrated that a modular data centre can be ready in 2-3 months. The site work started in the last week of November '23 for preparing the concrete base for the placement of the HPE Modular Data Centre (MDC) technology, known as a Performance Optimised Datacentre (POD) [21]. The model that was initially designed for Isambard 3's Grace Superchip technology was adapted to also house an HPE Cray EX 2500 liquid cooled cabinet, with 168 Grace-Hopper superchips (GH200) [19]. The GH200 superchip technology, combining NVIDIA's Grace ARM-based processor and H100 GPU, delivers a significant reduction in energy related to data movement as compared to an x86 plus H100 design. In less than 3 weeks, the DC20 was delivered on site, assembled, sealed, integrated with utilities such as power, cooling and networking, before being energised and powered up to rerun the factory Linpack/HPL top500 result on all 168 GH200s. These timelines are simply not feasible using a traditional bricks-and-mortar facility. Another significant advantage of the MDC technology is sustainability. The University of Bristol has an ambition to reach Net Zero by 2030. The MDC technology offers reuse of over 95% of the material used in the facility. Scope 1, 2 and 3 emissions throughout the lifecycle have been carefully recorded by the vendor in collaboration with their suppliers. Over 90% of the IT equipment will be decommissioned within the UK. Our target operational PUE is under 1.1.

The full Isambard-AI AIRR system will support a significant step-change in AI-enabled research in the UK [4]. Before Isambard-AI, the largest GPU-enabled systems were only a few hundred previous generation GPUs each. More recently, another AIRR resource called Dawn, consisting of around 1,000 Intel PVC GPUs, has come online in Cambridge, UK, and expected to go into production later in 2024 [16]. At 5,448 next-generation GPUs, and 21 ExaFLOP/s of 8-bit floating point performance for training LLMs, Isambard-AI represents about a 30X increase in performance for AI for the UK, over what was available in total in the country before. Isambard-AI represents an investment of over £200M in the hardware itself, including the new MDC and all plant (power, cooling etc). Including all the OPEX, which is mostly the power costs followed by the Isambard support team, Isambard-AI represents an investment of over £300M over the expected 5-6 years of the service.

Fig. 1 illustrates expected users and usage characteristics for AI platforms. Not only do these user communities need different scales of compute, from a single GPU up to thousands of devices, but also AI platform providers need to offer the look-and-feel of public cloud infrastructure-, platform- and software-as-a-service (IaaS, PaaS, and SaaS) interfaces. Interactivity, multi-tenancy, and lowering access barriers for diverse communities are among the key design patterns for supporting rapidly evolving needs for AI users that expect different Quality of Service (QoS) for their training and inference needs. These requirements are considerably different when compared with classic HPC simulation platforms.

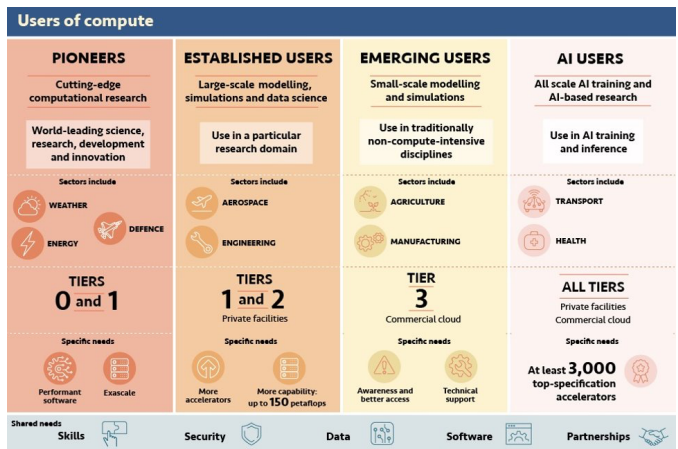

Fig. 1. An illustration from "The future of compute" report [3], comparing AI users with traditional HPC user communities.

In this paper we will describe the process of how we have been able to move so quickly – from first phone call about the idea, to lights on, within 12 months – in detail. We shall describe the full specification of the system, and detail the decisions behind the design, and how this is optimised for AI, including the differences compared to a traditional HPC system. We will also describe the software stack in detail, and explain how this will support AI research and users.

At the time of submitting the full manuscript (April 11, 2024), we only have NDA results collected by the vendors, which have been submitted for ISC24 Top500 [5]. Early AI (Artificial Intelligence) and ML (Machine Learning) benchmark results using MLPerf [6] have been gathered by the vendor on the phase 1 system. We have included early results for a set of AI and ML applications as well as HPC benchmarks on the system. The results validate our functional and performance expectations for a diverse set of applications.

Isambard-AI represents a step-change in what is possible. While many challenges remain, if successful, it will be one of the most rapid, large-scale supercomputer deployments ever, enabled by the MDC approach. Isambard-AI will be a new leadership-class system, delivering around 21 ExaFLOP/s of 8-bit performance for AI, and classifying as a significant pre-Exascale system (~250 PFLOP/s) in 64-bit. We will describe other noteworthy Isambard-AI features in this paper, including its class-leading energy efficiency, where the direct liquid cooling of the HPE EX4000s enables a PUE of under 1.1 during most operations [22]. Isambard-AI demonstrates that it should be possible to rapidly design, deploy and commission very large scale systems faster than ever before. The paper will share how we're achieving this in the hope that others will be able to reproduce our approach.

## II. BACKGROUND AND MOTIVATION

In Spring of 2023, an influential report was published in the UK titled "The Future of Compute" [3]. Several years in the making, this report included a number of important and influential recommendations, including one to establish a new AI Research Resource, or AIRR, for the UK. The report recognised the lack of AI-capable supercomputing facilities in the UK, and the impact this was having on AI research, as well as AI-enabled scientific discovery. In response, the UK Government announced a £900M fund to radically boost the UK's supercomputer capabilities. Supercomputer service providers were canvassed at the start of July 2023 to discover whether any could provide space, power and support for new, significantly-sized AI facilities in a much more rapid timeframe than usual: within 12 months. Thanks to already being in the process of procuring and building the Isambard 3 Tier-2 supercomputer in Bristol, an NVIDIA Grace-Grace CPU system of ~55,000 cores, in a new Modular Data Centre (MDC), the University of Bristol had already established it had a location with enough space and power (up to 5MW) to host a new AIRR system. Thus the University submitted a proposal for an Isambard-AI of 3,000 GPUs in mid-July 2023. One month later, on August 18th 2023, the Department of Science, Innovation and Technology (DSIT), the UK government branch driving the AIRR initiative, arranged the first call with the University of Bristol, and after a few weeks of rapid discussions, asked the Isambard team to design a system as large as possible, up to the 5MW power limit that the University had available at that time. The UK Government announced the award of the Isambard-AI project on September 13th 2023 [4]. The University then ran a rapid procurement process for a 5MW, AI-optimised solution, including an MDC solution, and the contract was awarded to HPE in mid-October. HPE and NVIDIA were able to fit 5,448 Grace-Hopper GPUs within Isambard-AI's 5MW limit, and thus the final system was significantly larger than the original proposal, at the Government's request.

The system is being delivered and is planned to be operational within 12 months of the contract being awarded; indeed, a Phase 1 system of 168 identical Grace-Hopper systems is delivered as an EX2500 within the Isambard 3 MDC, which itself arrived onsite in March 2024. Hence the first Isambard-AI users will be able to run early experiments on the Phase 1 system in the May/June 2024 timeframe, with Phase 1 planned for production in the summer of 2024. The site for the Phase 2 system, which includes 12 cabinets of EX4000 to deliver the remaining 5,280 Grace-Hopper GPUs, is due to be constructed next to the new Isambard 3 compound at the University of Bristol. The MDC approach enables this extremely rapid pace of development: the site is being designed and constructed within a University car park, the MDC is being delivered, and the Isambard-AI system then assembled and brought online, all within 12 months of contract signature, and in this case, within 15 months of the very first phone call to discuss the idea for the project. This rapid timeline also relies on the expertise of the Bristol team, which was already in the process of rapidly delivering the Isambard 3 system, at the same location and in another MDC, and thus the team had already built up experience of what was required to deliver systems as quickly as this.

## III. DESIGN SPECIFICATIONS

The Isambard-AI design criteria included rapid deployment, sustainability, Exascale AI performance, and cloud-native accessibility features for AI platforms and users. This section covers the specifications resulting from these design criteria, from data centre to multi-tenancy and elasticity, software stack to single-sign-on capabilities.

### A. Self Contained Units (SCUs) for rapid deployment and sustainability

The previous generations of Isambard 1 and 2 ARM based supercomputing platforms have been hosted at the UK Met Office (MO) in Exeter [27]. This option was no longer available for Isambard 3. Hence, we explored alternative solutions such as self-contained units (SCUs) for datacentre deployments that could be completed within 6 months, using the Modular Data Centre (MDC) solution. A car park at the National Composite Centre (NCC) on the edge of Bristol city, was already identified as a location with sufficient power, cooling, networking and heat reuse, and being on a science park, it also met the criteria of having rapid turnaround for planning permission applications (just 8 weeks, compared to at least 6 months in the centre of the city). A critical requirement for the solution was meeting the university's target of Net Zero by 2030, such that the scope 1, 2 and 3 emissions are monitored; the recyclability of the entire solution, including the data centre was also important. The target PUE was set to under 1.1.

### B. AI and performance optimised compute and network

The key requirements for the solution highlighted in the future of compute report (Fig. 1) was availability of at least 3,000 AI optimised compute units with mature software stacks for rapidly evolving AI frameworks, including large language models (LLMs), underlying platforms such as TensorFlow and PyTorch, and an ecosystem from low level libraries to software as a service (SaaS) options for a diverse community of users who are training models and running inference applications. These AI applications and frameworks require very high network bandwidth connections. A high performance computing fabric with RDMA support, standard Ethernet protocol support, and native MPI performance and scaling was required as part of the solution.

### C. Physical space limits

The space within the NCC car park available for the Isambard-AI data centre was limited, with approximately 550sqm (15.6m x 35m) available for deploying over 5,000 GPUs, including all necessary utilities such as transformers, chillers, pump skids and physical security infrastructure. Thus a very dense solution was required.

### D. Direct liquid cooling

Since the AI compute is largely provisioned with Graphical Processing Units (GPUs), each of which requires upwards of 500W of power and cooling, direct liquid cooling (DLC) was required for the higher energy efficiencies this approach enables. In fact, DLC was preferred for the majority of the IT equipment within the proposed solution.

### E. Storage and data platforms for AI

AI/ML payloads will require heterogenous storage solutions optimised for the performance requirements of training and inferencing, and for software defined storage (SDS) functionality to support multi-tenancy for the sensitive data across use cases from areas such as industrial, healthcare and medical. Training runs require high speed dataset reading and checkpoint writing. Inferencing runs need high read access rates with parallel data delivery paths to the processors. Fast response times for large models with billions of parameters is crucial. These requirements are somewhat different to HPC simulations, which are often served at scale by parallel file systems such as Lustre and GPFS. For high throughput read operations or high IOPs that are characteristics of AI payloads, a shared parallel file system alone is not sufficient. To solve this problem, we included two sets of requirements for storage, one for performance, and another for Quality of Service (QoS) management. For performance, we required a file system that can support both high bandwidth and high IOPs. For the QoS, an SDS solution was needed that can support different protocols for different AI platforms and applications.

In addition, the national AI compute facility will need to support the movement of data from different storage solutions to other national facilities, archiving on standard tape libraries, and to public clouds' object storage. Hence, a programmable, API and policy driven data motion solution needs be part of the storage and data platform design space for AI.

### F. High performance, cloud-native software ecosystem for AI platforms

As identified in the Future of Compute report [3], the system should be accessible to a wide range of users that are familiar with interactive access (Jupyter Notebooks enabled by Jupyter Hub services), Python e.g. Anaconda virtual environments, and running modern software pipelines involving containerised packaging and distribution technologies, as well as ssh access. The system is designed to support at least two classes of users. The first class of user will be running datasets of all scales. The second class of users will run on a cloud-like tenant for deploying AI and ML frameworks in containers or via interactive interfaces such as Jupyter Notebooks. Cloud-Native, as defined by the Cloud Native Computing Foundation (CNCF) [15], refers to a paradigm for deploying services that leverage cloud infrastructure capabilities to manage services at scale while reducing manual effort. A cloud native architecture typically leverages technologies such as infrastructure-as-code (IaC), microservices, containerization, continuous integration and continuous delivery (CI/CD), and DevOps. These practices maximise the benefits enabled by the fast-paced AI and ML applications requirements. As a result, these applications and frameworks are largely deployed in public cloud platforms and infrastructure because these capabilities are readily available there, unlike classic supercomputing environments. These features enable engineering teams to deliver software quickly and efficiently to keep pace with the rapidly evolving software and system stacks for AI and ML frameworks. Moreover, multi-tenancy and elasticity are included as key requirements to manage diverse infrastructure-, platform- and software-as-a-service (IaaS, PaaS, and SaaS). Both HPC style batch and interactive access patterns are considered as first-class citizens on Isambard-AI.

*G. Lowering accessibility barriers and federation*

According to the Future of Compute report, AI users cover a wide spectrum of users and use cases (Fig. 1). Lowering accessibility barriers via single sign-on, is one of the important requirements for Isambard-AI. Support for identity federation has been included for distributed AI and HPC workflows that span multiple Digital Research Infrastructures (DRIs) including AI RR.

*H. Cybersecurity and RBAC (roles based access controls)*

UK legislation requires a comprehensive view of security covering physical, environmental, technical and operational concerns, which have to be applied for Isambard-AI as a national AI RR facility. Therefore, from the ground up, an overarching security architecture is needed to ensure that the system is as secure as possible, by design, against the defined threat models. Key focus areas include mitigating external threats, since the system will be available to the UK academic and research communities. Regular security updates, controls, and penetration testing should be carried out on the systems to identity potential vulnerabilities and ensure these are treated for the entire ecosystem. Of particular importance to support IaaS, PaaS, and SaaS user and administrator roles on Isambard AI, roles based access controls (RBAC) with strong authentication (multi-factor) and limited duration tokens must be supported across the IT solutions. Moreover, the solutions should be open source and open standards based to interoperate with UKRI DRI cybersecurity task force recommendations. In short, the approach to security should provide protection against threats of sponsored hackers and corporate espionage, since this is a high-profile national AI RR facility.

IV. IMPLEMENTATION DETAILS AND EARLY RESULTS

This section provide solution and implementation details for fulfilling the design specifications for Isambard-AI.

*A. Self Contained Units (SCUs) for rapid deployment and sustainability*

Isambard-AI phase 1 (168 Grace-Hopper superchips [19]) and Isambard 3 (368 Grace-Grace superchips [20]) are housed in a next-generation HPE Performance Optimised Data centre (POD) to fulfil a complete "Self-Contained Unit (SCU)" requirement, including all necessary power and cooling infrastructure. A POD comprises an IT module (DC20), that includes a row of IT racks and a hot corridor (supercomputer equipment including servers, networking, and storage), associated power modules to supply power including UPS, a heat removal plant module, a pump skid module and an internal cooling Chilled Water (CW) system. The initial solution for the Grace-Grace superchip-based Isambard 3 offers up to 55kW/rack efficiencies to enable higher power distribution and cooling requirements when using the very latest CPU/GPU technologies. This POD was modified to house the Grace-Hopper EX2500 DLC cabinet. The POD also provides the capability and output for waste heat re-use all year round at the National Composite Centre (NCC) site in Bristol, with the ability to connect into 3rd party sources via a heat pump (for instance other buildings and facilities for general heating or technological demand). The heat removal plant and chillers have been designed to support the cooling capacity of up to 711 kW, with liquid temperatures of 23/15°C, an ambient temperature of 27°C DB and maximum free-cooling capacity. In total, two independent free-cooling chillers with over-dimensioned free-cooling capacity to achieve excellent energy efficiency are included in a parallel configuration. The heat removal plant supports the IT payload and is extensible, being ready for connection to a heat recovery plant. The mechanical system design assumes external air temperature at 27°C DB (Dry Bulb Temperature) / 18.3° WB (Wet Bulb Temperature), which covers nearly 100% (99.6% calculated) of the location weather conditions according to ASHRAE weather data.

The flexibility and rapid deployment of the MDC for Isambard-AI phase 1 has already been achieved. On March 11th 2024 the DC20 POD was delivered on trucks in parts that needed assembling on site. The assembly took just 24 hours, and IT equipment started to be installed the very next day. It took just two weeks to go from an empty concrete pad on the date the MDC arrived, to powering up the EX2500 supercomputer on March 26th and successfully re-running the factory tests, including HPL, across all 168 GH200 GPUs. During this timeframe a number of activities happened in parallel, since we were installing the new MDC-based data centre for the first time. This include the connection of utilities, such as power, cooling, network connectivity and security. The DC20 was assembled, pressure tested, and validated. The multi-step energisation process was completed safely to power up the Isambard-AI phase 1 supercomputer with 168 GH200 superchips in less than 2 weeks from the delivery of the POD itself.

Fig. 2 shows an overview of the Isambard compound at the NCC. Isambard-AI phase 1 and 3 POD deployment and integration is complete. The Isambard-AI phase 2 groundwork has since begun.

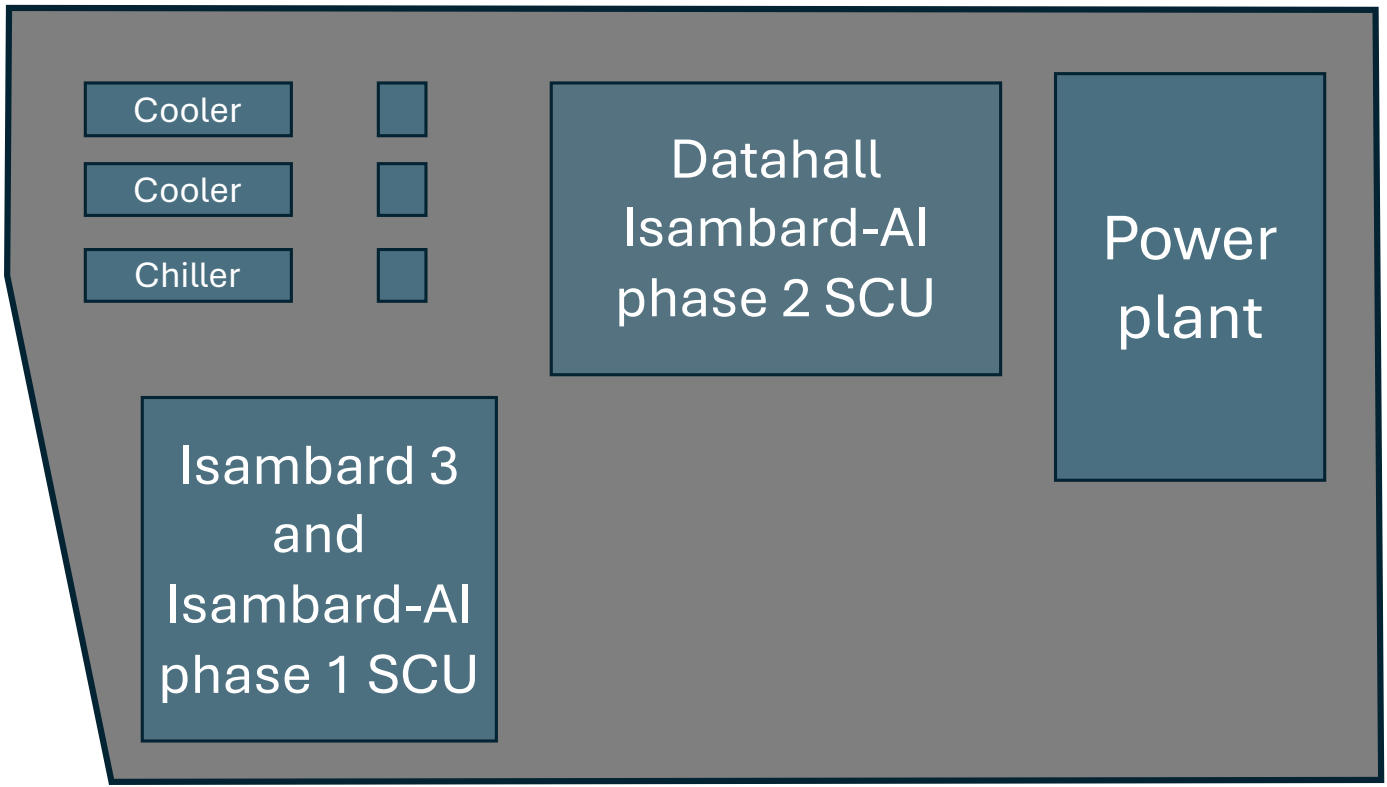

Fig. 2. A schematic of the Isambard-AI compound at the NCC, showing two Self Contained Units (SCUs), one for Isambard 3 and Isambard-AI phase 1 (deployed in March 2024) and another for Isambard-AI phase 2 scale out, arriving in summer 2024.

The MDC POD solution integrates a Data Centre Infrastructure Management (DCIM) system [7]. DCIM systems bridge the gap between IT and facility management of data centres by creating an IT and facility asset database. The database is directly linked to space, power, and cooling on the facility side, and to compute, storage, and networking on the IT side. Historically, DCIMs were targeted towards reliability and capacity planning; however, recently these systems have become indispensable for sustainability. We plan on capturing and correlating operational sustainability data with DCIM and the IT provisioning solutions for supercomputing ecosystems encompassing all Isambard MDCs and DRIs.

An additional benefit of the MDC is recyclability of the entire data centre, which is estimated to be over 90% of the material. It is becoming common practice to reuse and repurpose MDCs after 5-10 years of operations. As part of the university's Net Zero requirement, HPE projected scope 1, 2 and 3 emissions. Scope 1 is expected to be less than 1% because there is only a small diesel generator onsite for backup power during short-term power cuts (uncommon in our area). Around 90% of emissions will be scope 2 for running around 5 MW of supercomputing services. It is anticipated that the majority of the operations (over 95%) in Bristol will be performed with free air cooling, due to the moderate weather patterns in the region. Scope 3 emissions from the supply chain are estimated to be around 10%. As part of the contractual agreement, HPE will recycle over 90% of the IT equipment within the UK after decommissioning.

### B. AI and performance optimised compute and network

At the time of the procurement in Q3 2023, the most capable GPUs with a mature software stack for AI were NVIDIA's Hopper H100. The tensor cores within the H100 are optimised for AI floating point operations (FLOPS), including 8-bit tensor operations delivering approximately 4 PetaFLOPS per GPU. The Grace-Hopper Superchip, GH200, was expected to be not only a high performance combination of CPU and GPU compute and memory bandwidth, but also a highly energy efficient solution by reducing the cost of data movement between the CPU and GPU relative to a classic x86 and H100 architecture [19]. Fig 3 shows a single GH200 Superchip with Grace CPU LPDDR5 memory, H100 GPU HBM3 memory and NVlink Chip to Chip (C2C) bandwidth.

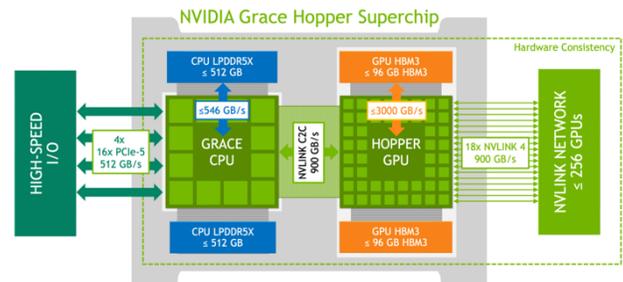

Fig. 3. Nvidia Grace-Hopper Superchip demonstrating memory and I/O bandwidth for the Grace CPU, Hopper GPU, NVlink C2C and I/O bandwidths. Source: Nvidia.

AI applications such as Large Language Models (LLMs) often require high aggregate memory capacity and bandwidth as well as high network injection and global bandwidth [29]. The node design for the Isambard-AI solution has four GH200 superchips connected via NVLink, as shown in Fig. 4, where each superchip has its own HPE Cray SlingShot network card capable of 200 Gbps [28]. Put together, AI applications running on Isambard-AI nodes can have about 850 GB of single memory address space, and order of 15 AI PetaFLOPs and 800 Gbps network injection, per node (4x GH200). The memory bandwidth available between each CPU and GPU is about 900 GB/s, using NVIDIA's Chip-to-Chip (NVLink C2C).

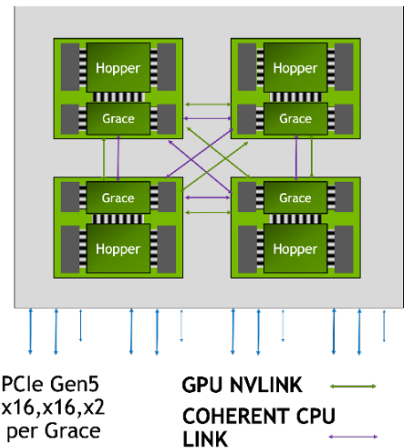

Fig. 4. Isambard-AI compute node with 4 GH200 Superchips. Source: HPE.

Isambard 3's node design, based on the Grace-Grace superchip, is different from Isambard-AI, as this is so far not offered as a dense HPE Cray EX, direct liquid cooled solution.

TABLE I. presents an overview of the three Isambard supercomputing platforms that are being installed in 2024. These include Isambard-AI phase 1 and phase 2 platforms with Grace-Hopper GH200 superchips, and the Isambard 3 Grace-Grace superchip and multi-architecture comparison system (MACS) nodes.

*C. Physical space limits*

The available space in the NCC car park for the MDC and all power and cooling ecosystem for Isambard-AI is 550sqm (15.6m x 35m). In order to deploy over 5,000 AI optimised GPUs, each possibly consuming over 600 Watts, within these constraints, we required a highly dense computing solution, such as HPE's Cray EX cabinets (custom sized). 12 computing cabinets, containing 5,280 Grace-Hopper GPUs (440 per cabinet) will be deployed, including the high performance networking infrastructure. As a comparison, an NVIDIA GH200 NVL32 rack-scale solution within NVIDIA DGX Cloud on AWS, contains 32 GH200 superchips per cabinet, connected with NVSwitch. Therefore a DGX GH200 SuperPod, defined as an AI Supercomputer, contains 256 GPUs in 16 cabinets. HPE's Cray EX 4000 cabinets offered an optimal solution for physical constraints (space and power) for Isambard-AI, delivering an order of magnitude more GPUs in the same space.

*D. Direct liquid cooling*

The HPE Cray EX liquid-cooled cabinet design can support up to 400 KW of power. This includes compute blades (Fig. 4) and high performance networking switch blades. All components are cooled by liquid cooling loops that run throughout the compute infrastructure. Each EX cabinet has a series of Power Distribution Units (PDUs) and rectifiers, which convert incoming AC power into DC power for distribution to the individual compute and switch blades with appropriate DC voltage conversions. The overall cooling loop is a closed loop that originates in the CDU. One CDU can support up to four liquid-cooled cabinets. The CDU maintains the cooling liquid at the specified temperature and removes the heat via a heat transfer mechanism to data centre water. The CDU requires an inlet water temperature of up to 32˚C, which helps eliminate the need for chillers, especially in Bristol's temperate climate, and further lowers energy usage. It is anticipated that the chillers will not be needed for 98% of operations. Cold plates remove heat from the compute units directly. The network mezzanine cards are cooled by the same compute cold plates. Connections to and from the compute and switch blades are quick connect and dripless, and allow a blade to be removed for servicing without shutting down the entire system. We expect the overall operational budget to be significantly lower than a similar sized air-cooled installation, with PUE efficiencies of < 1.1. Direct liquid cooling also increases the opportunity for, and effectiveness of, reusing the waste heat to heat local homes and businesses, an opportunity we intend to pursue in the future.

*E. Storage and data platforms for AI*

To address the diverse and rapidly evolving storage requirements for AI payloads, three different options have been included in the solution:

- An all-flash HPE ClusterStor E1000 based on Lustre. The scale-out Isambard-AI phase 2 will contain 44 flash storage servers with a total of 1,056 x 30.72TB NVMe SSDs, providing around 20.3 PiB of useable space. The aggregated IOR throughput is expected to be up to 1,980 GB/s for writes, and up to 2,500 GB/s for reads. The system can support 35 Million 4KB random read IOPS and 3.7M initial 4KB write IOPS.

- Local, 3.84 TB SSDs are included in a subset of compute nodes to support node local, small scale or sensitive payloads.

TABLE I. HIGH LEVEL TECHNICAL DETAILS OF ISAMBARD SUPSERCOMPUTING PLATFORM

|  | Isambard-AI Phase 1 | Isambard-AI phase 2 | Isambard 3 |
| --- | --- | --- | --- |

| Total nodes & superchips | 42 nodes / 168 NVIDIA GH200 superchips | 1,320 nodes / 5,280 NVIDIA GH200 superchips | 384 nodes / 384 NVIDIA Grace superchips |
|---|---|---|---|
| Compute per node | 288 Grace cores, 4 H100 GPUs | 288 Grace cores, 4 H100 GPUs | 144 Arm Neoverse v2 cores |
| Memory per node | 512 GB LPDDR + 384 GB HBM3 | 512 GB LPDDR5 + 384 GB HBM3 | 256 GB LPDDR5 |
| Network injection per node | 4 x 200 Gb/s Slingshot | 4 x 200 Gb/s Slingshot | 1 x 200 Gb/s Slingshot |
| Storage | ~1 PB all-flash ClusterStor E1000 Lustre | ~20.3 PB all-flash ClusterStor E1000 Lustre, ~3.56 PB VAST | ~2.3 PB ClusterStor E1000 Lustre |
| Programming environment | Cray PE / NVIDIA HPC SDK / Arm compilers | Cray PE / NVIDIA HPC SDK / Arm compilers | Cray PE / NVIDIA HPC SDK / Arm compilers |
| Types of compute racks | Cray EX2500 | Cray EX4000 | Cray XD2000 |
| Number of compute cabinets | 1 | 12 | 6 |
| MDC POD | DC20 (modified for EX2500) | DC10 with air cooled parts | DC20 air cooled |

- The multi-protocol and multi-tenant SDS solution is based on the VAST Data Platform appliance, consisting of 4 x protocol server enclosures (termed C-Boxes) and 3 x flash storage enclosures (termed D-Boxes) [18]. Each C-Box enclosure contains 4 x blade servers (termed C-Nodes) making 16 C-Nodes. The solution provides 3.56PB native useable capacity (before data reduction) and 5.6PB (5 PiB) logical capacity @1.6:1 reduction rate or 7PB logical capacity @2:1 with the VAST Cluster's data compression and deduplication mechanisms and similarity-based data reduction algorithms. This option also offers deployment of object storage.

A programmable, API and policy driven tool from HPE, called the Data Management Framework (DMF), will be used in co-designing data motion platforms for AI. These will support movement and staging of data from different storage solutions to other national facilities, archiving on standard tape libraries, and to public clouds' object storage. The tool supports roles-based access control (RBAC), and has native support for larger capacity parallel file systems such as Lustre, and cloud native object storage for supporting AI/HPC workflows. DMF supports traditional Hierarchical Storage Management (HSM) with incremental backups, file versioning as well as end-to-end data management for long term data retention. A DMF S3 mover supports data movement to public cloud object storage. A DMF tape mover will enable data movement to a standard tape library. Community-specific AI data platforms will be developed using the DMF APIs using open source and open standard solutions, such as S3 object store. This is part of the future work for phase 2, which will take place in conjunction with the VAST SDS solution.

### F. High performance cloud-native software ecosystem for AI platforms

AI and ML platforms often make use of popular cloud-native technologies to keep pace with rapid development and deployment cycles, often described as DevOps. In the case of Isambard-AI, we require the application of DevOps tools and practices for HPC. Cloud-like platforms usually treat multi-tenancy as a first-principles capability to align with modern DevOps management techniques. This in turn supports resource elasticity, and enables customers to deliver their own, custom IaaS, PaaS, and SaaS solutions efficiently in a reproducible manner. Classic HPC systems typically have a single configuration that is updated a maximum of a couple of times a year. Isambard-AI will be provisioned using HPE's Cray Systems Management (CSM), which is a cloud-native, Kubernetes-based, API-driven HPC systems software solution [24]. Different services layers are shown in Fig. 5.

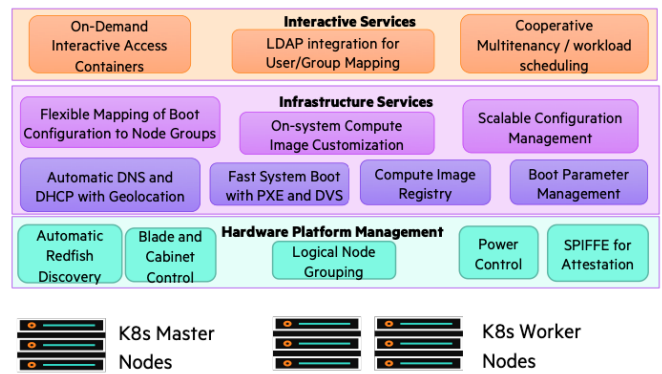

Fig. 5. A service level view of the CSM architecture. Source: HPE

CSM is open-source software (MIT), first released as closed source approximately four years ago, and has recently seen its sixth major release (1.5). We are using CSM as a foundation as it natively provides a software-defined, multi-tenancy architecture, supporting infrastructure-as-code (IaC) and DevOps approaches [26]. The concept of a multi-tenancy in this environment is managed by a tenancy "controller hub," called the Tenant and Partition Management System (TAPMS) [24]. This in turn builds upon the features in CSM, thereby inheriting the availability, scale, resiliency, disaster recovery, and security properties of the platform. CSM supports HPE Cray EX hardware and provides a base platform for several hardware and software product streams comprising a complete HPC solution, such as the Cray Operating System (COS), the System Monitoring Application (SMA), the Cray Programming Environment (CPE), and the Slingshot Highspeed Network (HSN) Fabric Manager (FM). CSM represents a notable paradigm shift in systems architecture, and is being successfully hardened by the first cohort of supercomputers it has been deployed to manage.

CSM heavily leverages Kubernetes (K8s), along with a larger portfolio of Cloud Native Compute Foundation Landscape and other open technologies. Management services in CSM run as micro-services inside K8s, using Istio for secure service-to-service communication, and builds upon the zero downtime features rooted in K8s. CSM further leverages KeyCloak (for federated identities) and Spire/SPIFFE (for non-person identities) to provide horizontally scaled, Zero Trust Architecture focused API and UI ingress services. CSM uses Ceph for remote block, clustered file systems, and S3-compatible utility storage.

Presently, all deployed services in the management and compute zones execute on bare metal, except for an emerging feature that uses virtualization techniques to emulate ARM OS image builds on x86_64 K8s worker nodes. An investigation is also currently underway to virtualize K8s master and worker nodes towards online migration between management clusters, and support for "blue green" promotion techniques. The objective is to improve the classic single-tenancy user supercomputing experience. CSM version 1.5 multi-tenancy is focused on PaaS-aligned, soft multi-tenancy with a plan to iterate towards "harder" multi-tenancy through key isolation capabilities. Resources owned by a "tenant" presently are bare-metal compute nodes and their software configuration. The IaC practices are aligned with DevOps and GitOps patterns. To support role-based tenant-driven configurability, CSM IAM (Identity and Access Management) persona have been updated to incorporate roles of a tenant administrator (least privilege perspective), and in addition to the existing global administrative persona as infrastructure administrator. The architectural guidelines have been updated for the development of the access control model of CSM providing an API to the HPE Cray EX supercomputer to configure the infrastructure dynamically. This in turn can leverage existing cloud infrastructure, modern software development that has shifted toward microservice architecture to increase scalability. It also reduces the complexity of services and applications. Details and different use cases are available in [23][24][25][26]. Fig. 6 illustrate the setup that we are leveraging for the Isambard-AI phase 1 multi-tenancy configuration.

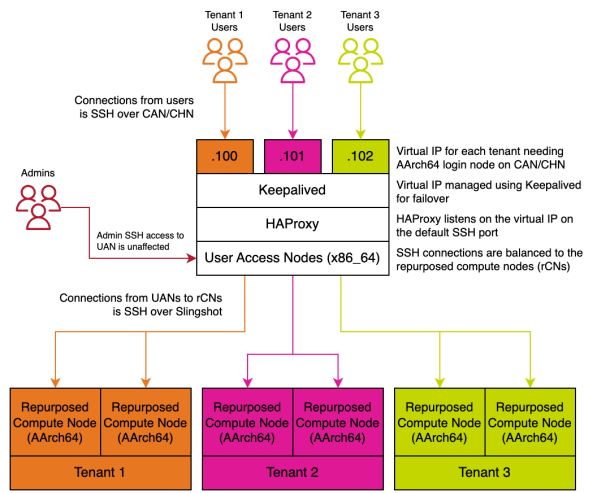

Fig. 6. Multi-tenant configuration for the Isambard-AI phase 1 system, utilising the x86-64 User Access Nodes as transparent proxies to compute nodes, repurposed as AArch64 Tenant login nodes, which are placed under the control of tenant administrators.

The CPUs in the Isambard-AI compute nodes are based on Grace-Hopper AArch64, i.e. Arm, while the Instruction Set Architecture (ISA) and the User Access Nodes are x86_64. It is preferable to present a single processor architecture to both users and tenant administrators. To this end, tenants are provisioned with a repurposed AArch64 compute node as a login or user access node. These "repurposed compute nodes" (rCNs) can be presented by tenant administrators to their users to use for a variety of purposes, e.g. running short interactive data analysis workflows, for local compilation of applications, or for running the front-end services for JupyterHub. Isambard-AI will initially present a JupyterHub tenant where the front-end JupyterHub services will run on one of these rCNs. Following a model deployed at the Swiss National Supercomputing Centre (CSCS), this will be backed by a Slurm queue that is also accessible from this rCN-fronted tenancy. A Slurm job will be automatically submitted when a user requests a new notebook. Once allocated to one of the compute nodes, this job will start a Jupyter notebook worker process, which is routed and presented to the user via the JupyterHub front-end running on the rCN.

Slurm is used as the resource manager in the first instance for reasons of speed of implementation. However, the goal is that CSM's support for Kubernetes for resource management will be deployed in a future iteration of the Jupyter notebook service. Using Kubernetes should support more fine-grained allocation of Jupyter notebook kernels to compute nodes (or parts of a compute node). Jupyter notebook workers could be scheduled into pods which are limited by the resources programmatically controlled via the container runtime. This is important, as compute nodes on Isambard-AI are large – 288 CPU cores, 4 H100s, 896 GB of total memory, and 4 200Gbps Slingshot injection points. Having mechanisms that enable multiple users to be partitioned from each other when sharing a node is necessary, particularly as the service should be welcoming to all users, regardless of whether they only need a single GPU for training for 1 hour, or thousands of GPUs for training for months.

A further benefit of using Kubernetes is that it will simplify the deployment of more complex MLOps-style workflow engines. Tools, such as MLFlow [8] and KubeFlow [9] involve multiple services running for training, experiment tracking, pipelines and dashboards. Kubernetes is becoming the ML community's workload resource manager of choice, so presenting Kubernetes as an option for tenant administrators is a primary goal of the Isambard-AI service.

Another distinguishing feature for supporting AI platforms as compared to classic simulation HPC platforms is the capacity and resource management and scheduling requirements. Broadly, the resource manager and scheduler need to support at least four common usage patterns for AI and ML applications and frameworks:

1. Experimentation: these are typically shorter period jobs with large capacity, interactivity and steering needs.
2. Training: these jobs are for longer periods, from days to months on large capacities.
3. Fine tuning: these are shorter period jobs with low capacity.
4. Online and offline inference: these depend on the usage patterns and can be served for implementing online and offline pipelines.

We leverage Google's AI computing or hypercomputer approaches for resource management and scheduling for AI and AI and HPC workflows [17]. These include:

- Flex Start with guaranteed completion, aka HPC/slurm batch—with periodic checkpoint/restart (2, 4).
- Calendar with start/stop time, aka HPC/slurm reservation—needs automation (1, 3 & 4)

Both modes will be implemented in a programmable manner on Isambard-AI with adaptations to the existing resource scheduler. We believe the resource management and scheduling patterns are simpler to implement for 1 to 1000s of GPUs on the tightly integrated Isambard-AI computer, because of its tightly-coupled high performance network and storage resources. This is unlike some public clouds, where it can be challenging to identify dedicated, tightly integrated resources within a single data centre environment.

### G. Lowering accessibility barriers and federation

Isambard-AI is designed to be welcoming to all users. Many in the AI community are new to supercomputing, and instead owe most of their experience to cloud-based AI services such as Google Colab [10] and Weights and Biases[1]. These services prioritise simple web-based portals backed by modern web-based logins and account management systems. To provide a similar experience for Isambard-AI, users will connect via a simple, web-based user access portal based on Waldur [11]. All resources will be provisionable and available via this portal. For example, the aim will be to both launch a Jupyter notebook session, and then directly connect to that session via web links provided through this portal. In addition, federated identity and access management will be provided via integration with MyAccessID [12] via eduGAIN [13] and its UK implementation via the UK Access Management Federation [14]. This federation enables users to identify themselves to the Waldur portal using credentials supplied by their institutional identity provider (IdP). Users connecting to the portal are redirected to their institutional login page of their local IdP to identify themselves via their normal processes (e.g. logging in using their institutional identity and password, and using their existing authenticator / multifactor authentication apps where needed). Upon successful completion, their name, institution and institutional email address are communicated via MyAccessID to Waldur as a set of login attributes with their access token. These attributes are used to access the user's account on Waldur if it exists, or to create a new account, and then direct the user to read and accept the terms and conditions of the Isambard-AI service. All user authentication is then handled via the user's Waldur account. This includes SSH logins, where a similar MyAccessID-backed user identification workflow will be used to generate SSH certificates signed by a Certificate Authority (CA) trusted by the login infrastructure of Isambard-AI.

### H. Cybersecurity and RBAC (roles based access controls)

The security architecture in an AI environment presents a different set of challenges to those in a 'traditional' enterprise data centre, which is also evolving from reliance upon host-based controls and tightly firewalled network segments to protect hosts. However, some of these approaches impact performance or impede technical innovation therefore aren't preferred in an HPC/AI data centre. The first part of the solution is the HPE Cray System Management (CSM) solution.

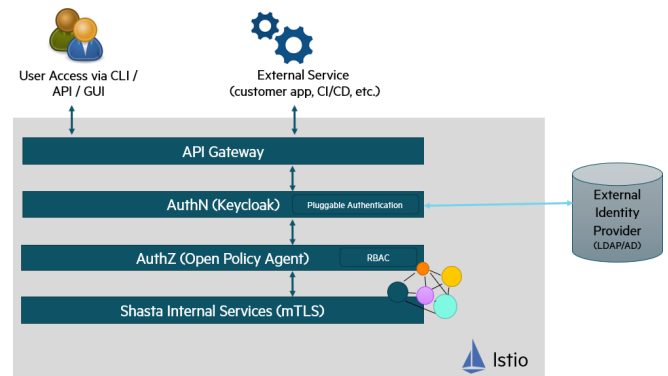

Fig. 7. Overview of CSM IAM

Access to all management services in CSM is via RESTful APIs and CLI commands. An API Gateway provides access to the HPE Cray EX System implementing identity and access management (IAM) for authentication and authorisation leveraging the capabilities of KeyCloak, Open Policy Agent (OPA), and Istio (Fig. 7). The IAM supports both human and non-human users with programmable RBAC interfaces. Access can be either via KeyCloak local accounts, or via connection to an external identity provider such as LDAP. RBAC are implemented by OPA. Istio provides mutual Transport Layer

---

[1] https://wandb.ai/site

Security (mTLS). Moreover, the solution offers multiple credential and certificate management options and there is auditable access to all APIs.

All management interaction requires TLS and strong authentication via RESTful APIs. Access to each API requires the presence of a valid, encrypted token that is authorised to perform a function. The narrowest permission structure permits access to a single REST verb on an API endpoint with a specific structure and set of query parameters providing highly granular control over the CLI for various management roles.

The Slingshot network provides security and isolation by constraining user space traffic to Virtual Networks: sets of endpoints that permit communication between them, but prevent access from outside of the virtual network. Each such virtual network is identified by an ID known as a Slingshot Virtual Network Identifier (VNI) that can be used for a tenant identifier in an overlay network. The network hardware enforces isolation between virtual networks using a VNI label carried in every RMA packet. The labels are not visible to or modifiable by the user.

System administration in ClusterStor is configured using role-based admin accounts. The Lustre servers are dedicated storage nodes, and are not accessed directly by users or administrators. Access is only possible via two dedicated ports on the System Management Units (EAN). Since no user codes are executed on the storage servers, an attack vector is unavailable.

*I. Early Performance Results*

The Isambard team had a few days to gather results using a diverse set of AI, ML and HPC benchmarks. These are out of the box measurements, therefore, we anticipate performance improvements after tuning. Nevertheless, the results confirm the design choice for 4xGH200 with SlingShot interconnect for a variety of benchmarks. Fig 8 shows single node results for a training benchmark with Slingshot 2.1, libfabric 1.15.2.0 and GPU RDMA enabled. These results are promising when compared to highly tuned, publicly reported data.

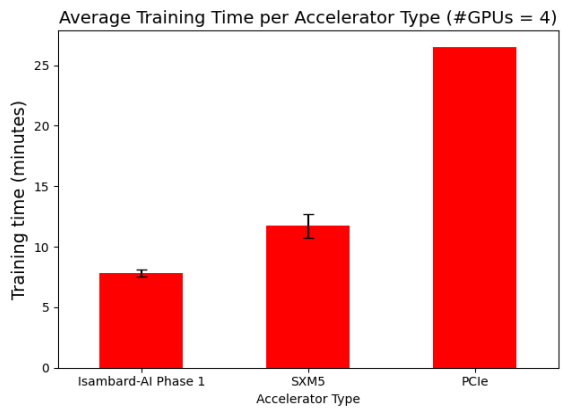

Fig. 8. MLPerf training with Bert-Large; lower times are better.

We gathered results for inference using llama.cpp (70 B) on the Grace CPU. Fig. 9 compares them with the publicly reported results from contemporary processors, even with a relatively low core and thread count per socket.

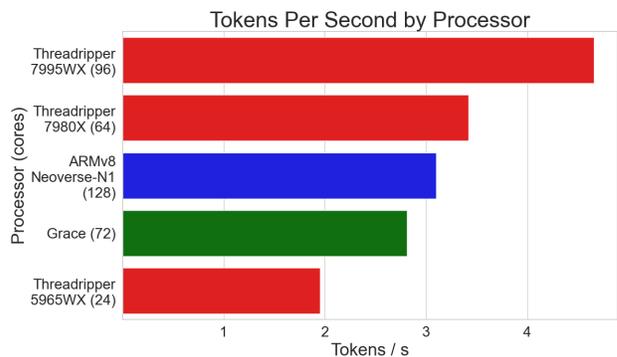

Fig. 9. llama.cpp (70 B) inference results; higher numbers are better.

The HPC capabilities of the Isambard-AI nodes and systems are measured using a set of synthetic microbenchmarks and mini-applications. Fig. 10 shows memory bandwidth results for multiple programming language implementations of the BabelStream benchmark. This includes CUDA, Thrust, Kokkos, OpenCL, SYCL (via AdaptiveCpp), C++ StdPar, OpenACC, Julia, and OpenMP target. Some of the compilers (GCC, Acpp, Clang) are built from source and everything worked as expected, achieving a high fraction of the peak memory bandwidth performance. This means that users can expect good out of the box performance, or performance portability, without undertaking significant tuning efforts.

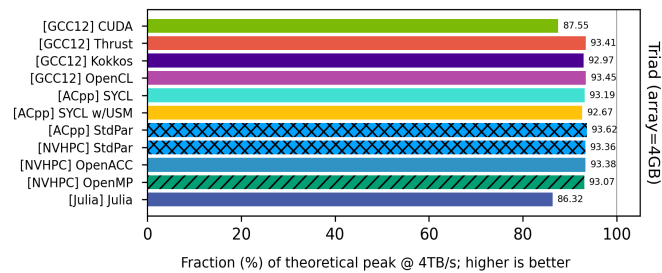

Fig. 10. https://github.com/UoB-HPC/BabelStream GH200 results.

CloverLeaf is a mini application that is part of SPEChpc2021. It is primarily memory bandwidth limited, based on a structured grid, stencil pattern. We use the CUDA version. Fig 11 shows weak scaling up to almost the entire machine of 160 GPUs (8 GPUs are on 2 repurposed login nodes). We observed similar results for TeaLeaf (also part of SPEChpc2021), which is also primarily memory bandwidth intensive with MPI collectives.

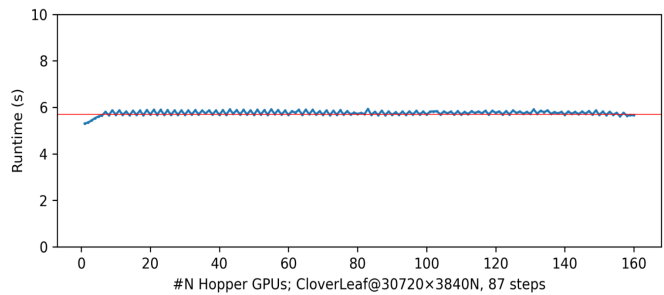

Fig. 11. CloverLeaf mini-app performance and scaling results.

## V. Future outlook

At the time of writing, the first Isambard Modular Data Centre is up and running, and the Isambard-AI phase 1 system of 168 Grace-Hopper GPUs is installed and passing its tests. The Isambard 3 CPU system, based on NVIDIA's Grace-Grace CPUs, is due to be installed in May 2024 in the same MDC. The scale-out Isambard-AI phase 2 system, of the remaining 5,280 GPUs, is due to be installed in a second, larger MDC solution, being constructed now, with operations due to commence by the end of the year. At this point, three separate but closely related new Top500 supercomputers will be co-located in the Isambard compound in Bristol, having been built from scratch (all new MDCs) in around 12 months. We believe this is the first time this has even been achieved.

### Acknowledgments

Isambard-AI is funded by the UK Government's Department of Science, Innovation and Technology (DSIT) via UKRI / STFC. The project has only been possible thanks to a very talented and extremely hardworking team of people across the University of Bristol, HPE, Contour, NVIDIA, the National Composites Centre, and our building contractors, Oakland. AI/ML and HPC early performance results were collected by the members of the Isambard team, Wahab Kawafi and Tom Lin, respectively.